\documentclass[12pt]{article}
\usepackage{psfig,cite}

\makeatletter
\renewcommand\section{\@startsection {section}{1}{\z@}%
                                   {-3.5ex \@plus -1ex \@minus -.2ex}%
                                   {2.3ex \@plus.2ex}%
                                   {\normalfont%\sffamily
                                     \large\bfseries}}
\makeatother

\newcommand{\ri}{\mbox{\footnotesize I}}

\date{}
\title{
{\large\rm Cavendish-HEP 98/03}\hfill{\large\rm March 1998}\\
{\large\rm DAMTP-98-14}\hfill
\vspace*{2.5cm}\\
Small-$x$ Parton Distributions\\ of Large Hadronic Targets
\\[1.5cm]}
\author{
A. Hebecker\\
{\normalsize\it DAMTP, Cambridge University, Cambridge CB3 9EW, England}
\\[.5cm]
and\\[.5cm]
H. Weigert\\
{\normalsize\it Cavendish Laboratory, Cambridge University, Cambridge CB3 0HE,
England}
\vspace*{2cm}\\
}
\begin{document}

\setlength{\baselineskip}{18pt}
\maketitle
\begin{abstract}
\noindent
A simple and intuitive calculation, based on the semiclassical
approximation, demonstrates how the large size of a hadronic target
introduces a new perturbative scale into the process of small-$x$ deep
inelastic scattering. The above calculation, which is performed in the
target rest frame, is compared to the McLerran-Venugopalan model for
scattering off large nuclei, which has first highlighted this effect
in the infinite momentum frame. It is shown that the two approaches,
i.e., the rest frame based semiclassical calculation and the infinite
momentum frame based McLerran-Venugopalan approach are quantitatively
consistent.
\end{abstract}
\setcounter{page}{0}
\thispagestyle{empty}
\newpage

\section{Introduction}
It has been suggested by McLerran and Venugopalan \cite{mv} that for
very large nuclei the parton distribution functions at small $x$ are
perturbatively computable. Their argument relies on the high density
of partons per unit area per unit rapidity which serves as a new hard
scale in the problem. Extending these ideas, an evolution equation has
been formulated \cite{JKWev} to follow the growth of this potentially
large density as $x$ decreases. These calculations are formulated in
the infinite momentum frame of the hadron.

In an apparently quite different approach, small-$x$ deep inelastic
scattering is described in the target rest frame, where the long-lived
partonic fluctuations of the very energetic photon scatter off the
target hadron \cite{bk}. Explicit formulae focusing on the $q\bar{q}$
component of the photon can be found in \cite{nz}. In the
semiclassical framework, the scattering of the partonic fluctuations
off the target is modeled by an eikonal interaction with the target
colour field \cite{nac,bal,bhm}.

In the present paper, the semiclassical approach in the target rest
frame is applied to the case of very large hadronic targets, such as
heavy nuclei.  The perturbative calculability of parton densities
advertised previously emerges in a very simple and intuitive way. It
is the result of the limited transverse size of the relevant
$q\bar{q}$ fluctuations of the virtual photon. For an ordinary
hadronic target, leading twist contributions to deep inelastic
scattering come from both small and large size $q\bar{q}$
configurations. For larger targets, the $q\bar{q}$ pair has to
penetrate an extended colour field and the geometric limit of the
cross section is already reached for small size pairs. Accordingly,
non-perturbative large size $q\bar{q}$ configurations do not
contribute significantly to the total photon hadron cross section.

We then go on to compare the semiclassical approach with the
McLerran-Ve\-nu\-go\-pa\-lan approach in greater detail.  It is shown
that they can be mapped onto each other once the non-Abelian eikonal
factors characteristic of the target colour field are recognised as
the basic building blocks of both methods. We confront the different
perspectives inherent to the rest frame and infinite momentum frame
calculation.

Performing an average over different colour field configurations,
explicit formulae are obtained for the dipole cross section
$\sigma(\varrho)$, which describes the interaction of a $q\bar{q}$
pair of transverse size $\varrho$ with the target.

Although the above results are only valid in the limit of very large
targets, we believe they are useful in a much wider sense. They
provide a theoretical laboratory in which, starting from a
non-perturbative yet technically manageable description of the target,
different small-$x$ processes can be investigated. In particular, they
establish a non-trivial limit in which the semiclassical approach can
be justified in the framework of QCD.

The paper is organized as follows. After summarising the target rest
frame view on deep inelastic scattering, a simple derivation of the
perturbative scale associated with large targets is given in
Sect.~\ref{trf}. The equivalence to the McLerran-Venugopalan approach
is demonstrated in Sect.~\ref{imf} and explicit models for the
averaging over the gluon field are described in Sect.~\ref{ems}. The
conclusions are given in Sect.~\ref{conc}.

\section{Calculation in the target rest frame}\label{trf}
\begin{figure}[ht]
\begin{center}
\parbox[b]{10cm}{\psfig{width=10cm,file=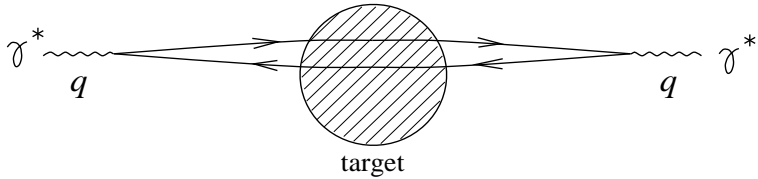}}\\
\end{center}
\caption{The Compton scattering amplitude within the 
semiclassical approach.
\label{comp}}
\end{figure}

The total cross section for the scattering of a virtual photon off a
hadronic target can be obtained from the Compton scattering amplitude
via the optical theorem. In the target rest frame, the leading order
partonic process is the fluctuation of the photon into a $q\bar{q}$
pair (see Fig.~\ref{comp}). To keep calculations as simple as
possible, we consider a longitudinally polarized photon coupled to
scalar quarks with one unit of electric charge. As far as the
$Q^2$-behaviour is concerned ($Q^2\!=\!  -q^2$), this is analogous to
the standard partonic process where a transverse photon couples to
spinor quarks \cite{bd}.

In analogy to \cite{nz}, the longitudinal cross section can be written
as
\begin{equation}
\sigma_L=\int d^2\varrho_\perp \sigma(\varrho)
  W_L(\varrho)\,,\label{conv}
\end{equation}
with the square of the wave function of the virtual photon given by 
\begin{equation}
W_L(\varrho)=\frac{3\alpha_{\mbox{\footnotesize em}}}
  {4\pi^2}\int d\alpha\,N^2
K_0^2(N\varrho)\,.\label{wl}
\end{equation}
Here $\varrho=|\varrho_\perp|$ is the transverse size of the
$q\bar{q}$ pair, $\alpha$ is the longitudinal momentum fraction of the
photon carried by the antiquark, $N^2=\alpha(1-\alpha)Q^2$, and $K_0$
is a modified Bessel function. Note that, in contrast to \cite{nz}, we
have defined $W_L$ to include the integration over $\alpha$.

The cross section for the realistic case of a transverse photon and
spinor quarks is obtained by substituting $K_0^2(\varrho N)$ with
$2[\alpha^2+(1-\alpha)^2]K_1^2(\varrho N)$ in Eq.~(\ref{wl}). None
of the qualitative results derived below is affected by this
substitution.

Within the semiclassical approach, the dipole cross section
$\sigma(\varrho)$ is given by
\begin{equation}
\sigma(\varrho)=\frac{2}{3}\int d^2x_\perp
\mbox{tr}\left[\mbox{\bf 1}-U(x_\perp)
U^\dagger(x_\perp+\varrho_\perp)\right]\,,\label{sc}
\end{equation}
where the SU(3) matrices $U$ and $U^\dagger$ represent the non-Abelian
eikonal factors associated with the quark and antiquark propagating
through the target colour field and averaging over all field
configurations is implicit.

\begin{figure}[ht]
\begin{center}
\parbox[b]{8cm}{\psfig{width=8cm,file=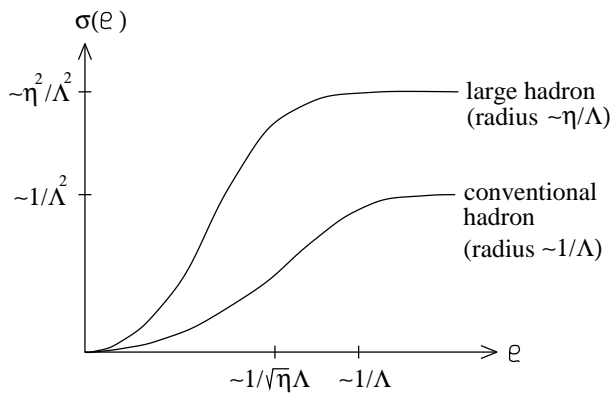}}\\
\end{center}
\caption{Qualitative behaviour of the function $\sigma(\varrho)$.
\label{sigma}}
\end{figure}

The functional form of $\sigma(\varrho)$ is shown qualitatively in
Fig.~\ref{sigma}. For conventional hadrons of size $\sim 1/\Lambda$
(where $\Lambda\sim\Lambda_{\mbox{\footnotesize QCD}}$) its typical
features are the quadratic rise at small $\varrho$
($\sigma(\varrho)\sim \varrho^2$ with a proportionality constant
${\cal O}(1)$) and the saturation at $\sigma(\varrho) \sim
1/\Lambda^2$, which occurs at $\varrho\sim 1/\Lambda$. Consider now
the idealized case of a very large target of size $\eta/\Lambda$ with
$\eta\gg 1$ ($\eta\sim A^{1/3}$ for a nucleus). It is easy
to see that at small $\varrho$ the functional
behaviour is given by $\sigma(\varrho)\sim\eta^3\varrho^2$ while
saturation has to occur at $\sigma(\varrho)\sim\eta^2/\Lambda^2$ for
geometrical reasons. It follows that the change from quadratic rise to
constant behaviour takes place at $\varrho \sim 1/\sqrt{\eta}\Lambda$,
i.e., at smaller $\varrho$ than for conventional targets.

{}From the above behaviour of $\sigma(\varrho)$ we will now derive the
dominance of small transverse distances in the convolution integral of
Eq.~(\ref{conv}). To do this, a better understanding of the function
$W_L(\varrho)$ is necessary. Recalling that $K_0(x)\sim\ln(1/x)$ for
$x\ll 1$ while being exponentially suppressed for $x\gg 1$, it is easy
to see that $W_L(\varrho)\sim Q^2\ln^2(1/\varrho^2Q^2)$ for
$\varrho\ll 1/Q$ and $W_L(\varrho)\sim 1/\varrho^4Q^2$ for $\varrho\gg
1/Q$. Here numerical constants and non-leading terms have been
suppressed.

Under the assumption $\Lambda^2\ll \eta\Lambda^2\ll Q^2$, the integral
in Eq.~(\ref{conv}) can now be estimated by decomposing it into three
regions with qualitatively different behaviour of the functions
$W_L(\varrho)$ and $\sigma(\varrho)$,
\begin{equation}
\sigma_L=\sigma_L^{\ri}+\sigma_L^{\ri\!\ri}+\sigma_L^{\ri\!\ri\!\ri}=\left(
\int_0^{1/Q^2}+\int_{1/Q^2}^{1/\eta \Lambda^2}+
\int_{1/\eta \Lambda^2}^{\infty}\right)\pi d\varrho^2
\sigma(\varrho)W_L(\varrho)\,.
\end{equation}
Of the three contributions 
\begin{eqnarray}
\sigma_L^{\ri}\qquad\sim &\displaystyle\int_0^{1/Q^2}
d\varrho^2\,\eta^3\varrho^2\,
\,Q^2\ln^2(1/\varrho^2Q^2)&\sim\quad\frac{\eta^3}{Q^2}
\nonumber \\ \nonumber\\
\sigma_L^{\ri\!\ri}\qquad\sim &
\displaystyle\int_{1/Q^2}^{1/\eta \Lambda^2}d
\varrho^2\,\eta^3\varrho^2\,\frac{1}{\varrho^4Q^2}&
\sim\quad\frac{\eta^3}{Q^2}\ln(Q^2/
\eta\Lambda^2)
\\ \nonumber\\
\sigma_L^{\ri\!\ri\!\ri}\qquad\sim &\displaystyle
\int_{1/\eta \Lambda^2}^{\infty}d\varrho^2\,\frac{\eta^2}
{\Lambda^2}\,\frac{1}
{\varrho^4Q^2}&\sim\quad\frac{\eta^3}{Q^2}\nonumber
\end{eqnarray}
the second one dominates, giving the total cross section 
\begin{equation}
\sigma_L\sim\frac{\eta^3}{Q^2}\ln(Q^2/\eta\Lambda^2)\,.
\end{equation}
It is crucial that the third integral is dominated by contributions
from its lower limit. Therefore, the overall result is not sensitive
to values of $\varrho$ that are larger than $1/\sqrt{\eta}\Lambda$. Phrased
differently, for sufficiently large targets the transverse size of the
$q\bar{q}$ component of the photon wave function stays perturbative.

\section{Calculation in the infinite momentum frame}\label{imf}
\begin{figure}[ht]
\begin{center}
\parbox[b]{6.3cm}{\psfig{width=6.3cm,file=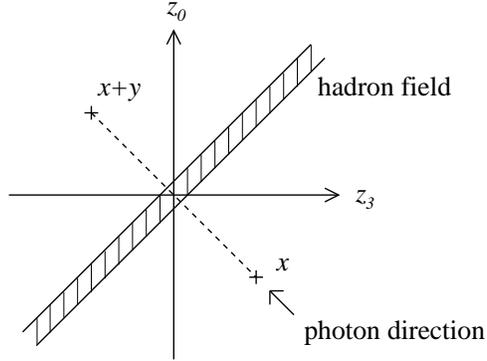}}\\
\end{center}
\caption{Space-time picture in the infinite momentum frame.
\label{st}}
\end{figure}

The above considerations can be related to the infinite momentum frame
calculation of parton distributions within the McLerran-Venugopalan
model \cite{mv1}. Here we use the term infinite momentum frame for a
reference frame where the `+'-component of the hadron momentum and the
`--'-component of the photon momentum are large ($k_\pm=(k_0\pm k_3)/
\sqrt{2}$). We start with the operator definition of scalar parton
distributions
\begin{equation}
q(\xi)=\frac{\xi P_+^2}{\pi}\int\! dy_-dx_-d^2x_\perp e^{-i\xi P_+y_-}
<\!P|\mathrm{T}\phi^\dagger(0,x_-+y_-,x_\perp)U_{x_-+y_-,x_-} 
\phi(0,x_-,x_\perp)|P\!>.
\label{qdef}
\end{equation}
Note, that we use a time ordered product in this definition (see e.g.
\cite{jaffe}) and choose to normalize the hadron state by demanding
$<\!P|P\!>=1$. The matrix $U$ is the standard link operator required
for gauge invariance of the above definition. The corresponding space
time picture for a hadron localized at $z_-=0$ is illustrated in
Fig.~\ref{st}.

It is understood that only the connected diagrams contribute to the
definition of $q(\xi)$ in Eq.~(\ref{qdef}), which is equivalent to
subtracting the vacuum expectation value of $\phi^\dagger\phi$.  It
follows that we just need to calculate the propagator of the scalar
field in the presence of the hadron and to subtract the propagator in
the vacuum,
\begin{equation}
q(\xi)=\frac{\xi P_+^2}{\pi}\int\! dy_-dx_-d^2x_\perp 
e^{-i\xi P_+y_-}
\mathrm{tr}\left[U(x_\perp)G(x,x+y)-G_0(x,x+y)\right]\,.
\label{qdef1}
\end{equation}
Here $x=(0,x_-,x_\perp)$, $y=(0,y_-,0)$ and $G_0$ is the conventional
Feynman propagator.

At very small $x$, the contribution of the constituent quarks to the
quark distribution is negligible. It is therefore sufficient to
calculate the propagator in the background of a certain colour field
and to average over all field configurations of the hadronic state. In
a gauge in which the potential vanishes outside the hadron (say,
outside the shaded region of Fig.~\ref{st}) the propagating particle
is simply rotated in colour space at the moment when it penetrates the
field. The high energy limit makes the region with non-zero potential
arbitrarily narrow in $z_-$-direction so that, for $y_->0$ and
$x_-<0$, the propagator can be written as
\begin{equation}
G(x,y)=\int\!\! d^4z\,G_0(x,z)
\left[2\delta(z_-)U^\dagger(z_\perp)\frac{\partial}
{i\partial z_+}\right]G_0(z,y)\,.\label{prop}
\end{equation}
Note that this corresponds to the propagation of a positive energy
antiquark from $x$ to $y$. The non-Abelian eikonal factor at transverse
position $z_\perp$ is explicitly given by
\begin{equation}
  \label{eq:repboost}
  U(z_\perp) = P \exp - i g \int_{-\infty}^{+\infty} d z_- 
  A_+(z_-,z_\perp)\,,
\end{equation}
where $A_\mu$ is the gauge potential of the hadron. An expression
analogous to Eq.~(\ref{prop}) holds in the region $y_-<0$ and $x_->0$,
while for the remaining regions where $x_-$ and $y_-$ are either both
positive or both negative the propagator is simply the free one.
This intuitive physical picture is slightly less apparent if one uses
an $A_+=0$ gauge as in \cite{mv1}. Then the propagators contain an
additional eikonal factor corresponding to the relative gauge
transformation while the link operator in Eq.~(\ref{qdef}) is absent.

The expression for the propagator provided by Eq.~(\ref{prop}) is exact
for a $\delta$-function-like source. Within the present context, the
highly Lorentz contracted target has a $\delta$-function-like
appearance on scales larger than $1/P_+$. The use of Eq.~(\ref{prop})
for the quark density calculation according to Eq.~(\ref{qdef1}) is
justified since typical values of $y_-\sim 1/\xi P_+$ are much larger
than $1/P_+$.

Note furthermore, that the gauge potential in Eq.~(\ref{eq:repboost})
is extremely large in the infinite momentum frame. However, this is
compensated by the smallness of the relevant integration region in
$z_-$. Clearly, the explicit formula for the eikonal factor
Eq.~(\ref{eq:repboost}) is invariant with respect to arbitrary boosts
along the $z_3$-axis.

Inserting our expression for the propagator into Eq.~(\ref{qdef1}) and
performing the necessary integrations \cite{mv1} the following result
is obtained \footnote{Note that a sign error in Eq.~(47) of \cite{mv1}
  leads to a spurious cancellation which affects their result.}
\begin{equation}
\xi q(\xi)=\frac{3}{(2\pi)^4}\int_0^{\Lambda_{UV}^2}N^2dN^2
\int d^2\varrho_\perp
K_0^2(N\varrho)\sigma(\varrho)\,.\label{qd}
\end{equation}
The dipole cross section $\sigma(\varrho)$ (see Eq.~(\ref{sc}))
emerges naturally due to the presence of two eikonal factors from the
combination of Eqs.~(\ref{qdef}) and~(\ref{prop}). The ultraviolet
cutoff $\Lambda_{UV}\ll Q$ is introduced to suppress the contribution
from partons with very high transverse momenta (compare the discussion
in \cite{bhq}).

Next, we want to explain in which sense Eq.~(\ref{qd}) is equivalent
to the cross section formula Eq.~(\ref{conv}). The cross section
$\sigma_L$ receives leading twist contributions from two regions, the
aligned jet region, where $\alpha\ll 1$ or $(1-\alpha)\ll 1$ and
$\varrho$ is large, and the high-$p_\perp$ region, where $\alpha\sim
1/2$ and $\varrho$ is small.  The small-$\alpha$ contribution of
Eq.~(\ref{conv}),
\begin{equation}
\sigma_{L,q}=\frac{3\alpha_{\mbox{\footnotesize em}}}{4\pi^2Q^2}
\int_0^{\Lambda_{UV}^2} dN^2\int d^2\varrho_\perp N^2
K_0^2(N\varrho)\sigma(\varrho)
\,,
\end{equation}
corresponds, in parton model language, to the contribution from the
quark distribution of the target (hence the index $q$). The latter can
be identified by comparing with the parton model formula (for scalar
partons)
\begin{equation}
\sigma_{L,q}=\frac{4\pi^2\alpha_{\mbox{\footnotesize em}}}{Q^2}xq(x)\,.
\end{equation}
As expected, the obtained quark distribution is in agreement with
Eq.~(\ref{qd}).

The perturbative nature of the scattering process off large nuclei, as
derived in Sect.~\ref{trf} from generic features of $\sigma(\varrho)$,
is implemented within the McLerran-Venugopalan model via the colour
field average implicit in Eq.~(\ref{qd}). This is subject of the next
section.

\section{Explicit models for $\sigma(\varrho)$}\label{ems}
Consider first a particularly simple model that can be analysed with
minimal calculational effort.

According to Eq.~(\ref{sc}), the expectation value of $UU^\dagger$ in
the background of a hadron is required. The averaging over the field
configurations of an ordinary hadron with radius $\sim 1/\Lambda$ is
symbolized by $\langle\cdots\rangle_{1/\Lambda}$. For values of
$x_\perp$ that correspond to a central collision and in the limit of
small $\varrho$, we have
\begin{equation}
\left\langle\,U_{ij}(x_\perp)U^\dagger_{kl}(x_\perp+\varrho_\perp)\,
\right\rangle_{1/\Lambda}\simeq \delta_{ij}\delta_{kl}-B\varrho^2
\Lambda^2
T^a_{ij}T^a_{kl}\,.\label{nh}
\end{equation}
Here $i,j,k,l$ are colour indices, the matrices $T^a$ form the
conventional basis of the SU(3) Lie algebra and $B$ is a constant. The
simple colour structure follows from the fact that, at small
$\varrho$, the path ordered exponentials implicit in $UU^\dagger$ can
be expanded in powers of the gauge field $A^a$.  The leading
contribution comes from $\langle A^aA^b\rangle\sim\delta^{ab}$.

To extend Eq.~(\ref{nh}) to the case of a large hadron, an additional
assumption has to be made. It is assumed that the volume of the hadron
can be split into mutually uncorrelated regions of size $\sim
1/\Lambda$, as the naive geometrical picture of a large nucleus built
from nucleons suggests.  If the linear extension of the hadron is
$\sim\eta/\Lambda$ and $n\simeq \eta$ is an integer, the $q\bar{q}$
pair passes approximately $n$ such regions on its way through the hadron
(see Fig.~\ref{comp}). The expectation value of $UU^\dagger$ is given by
\begin{equation} 
\left\langle\,U(x_\perp)U^\dagger(x_\perp+\varrho_\perp)\,
\right\rangle_{\eta/\Lambda}\simeq\left(\,\left\langle\,
    U(x_\perp)U^\dagger
(x_\perp+\varrho_\perp)\,\right\rangle_{1/\Lambda}\,\right)^n\,.
\label{prod}
\end{equation}
Here the appropriate contraction of the colour indices on the r.h.
side is implied. Using Eq.~(\ref{prod}) the trace of Eq.~(\ref{nh})
can be calculated for the large hadron case. For sufficiently large
$\eta$ and small $\varrho$, the product of $U$'s exponentiates giving
\begin{equation}
\left\langle\,\mbox{tr}\left[
    U(x_\perp)U^\dagger(x_\perp+\varrho_\perp)\right]\,
\right\rangle_{\eta/\Lambda}\simeq N_c
\left(1-C_F B\varrho^2\Lambda^2\right)^n
\simeq N_c\mbox{exp}[-\eta C_F B\varrho^2\Lambda^2]\,\label{expon}
\end{equation}
with $C_F=(N_c^2-1)/2N_c$.  Again, $x_\perp$ is chosen such as to
describe a central collision.  Therefore, the above is valid for an
$x_\perp$ area of size $\sim\eta^2/ \Lambda^2$. Inserting this into
Eq.~(\ref{sc}) the following result is obtained
\begin{equation}
\sigma(\varrho)\sim\frac{\eta^2}{\Lambda^2}
N_c\left(1-\mbox{exp}[-\eta C_F B\varrho^2
\Lambda^2]\right)\,.\label{m1}
\end{equation}
Clearly, this simple model is consistent with the generic features of
$\sigma(\varrho)$ discussed in Sect.~\ref{trf}. The above expansion in
powers of the gauge potential is only valid if $\varrho\ll 1/\Lambda$.
The exponentiation in Eq.~(\ref{expon}) is valid in the limit of large
$\eta$ for $\varrho\stackrel{<}{\sim}1/\sqrt{\eta}\Lambda$. Both
conditions are fulfilled in deep inelastic scattering in the limit of
very large hadronic targets. The simple formula of Eq.~(\ref{m1}),
which includes the numerical constant $B$ and an unknown overall
normalization, provides a model for $\sigma(\varrho)$ in this limit.

Let us now consider the same quantity within the framework of
\cite{mv,mv1}. Here, one of the central issues was to provide an
explicit prescription for the averaging over the target's colour
fields. The initial suggestion, the McLerran-Venugopalan model, used a
Gaussian statistical weight for the colour charge in the target that
generates these fields (compare also the discussion in
\cite{kovchegov}). Later, an evolution equation was derived to
determine such a statistical weight and its $x$ dependence
\cite{JKWev}. These results suggest that, although a Gaussian weight
is not a solution of the evolution equation, it may be used at not too
small $x$, as long as the parton densities are not too high. In the
present calculation we satisfy ourselves with the simple Gaussian
weight introduced in \cite{mv}.

As in~\cite{mv,JKWev}, we calculate the classical field for a static,
i.e. $x_+$ independent, colour source of the form
\begin{equation}
  \label{eq:source}
  J^a_\mu(x)  =  \rho^a(x_-,x_\perp)g_{\mu\, -}\,,
\end{equation}
where $\rho$ parametrises the adjoint charge density of the hadron
(not to be confused with the size $\varrho$ of the $q\bar q$ pair
used above). In the $A_-=0$ gauge the corresponding gauge field is
given explicitly by
\begin{equation}
  \label{eq:fields}
  \nabla_\perp^2A_+(x_-,x_\perp) = g \rho(x_-,x_\perp)\,;\qquad
  A_\perp(x_-,x_\perp)=0\,,
\end{equation}
where $g$ is the gauge coupling. 

Note, that here we are concerned with the space time region where the
potential in Eq.~(\ref{eq:repboost}) is nonzero. From the point of view
of Eq.~(\ref{prop}), where much larger distances in $x_-$ dominate,
this corresponds to resolving the structure of the $\delta$-function in
$x_-$.

According to \cite{mv}, the Gaussian average over $\rho$ is implemented
independently at each position in $x_-$. This means that the average of
an arbitrary functional $O[\rho]$ is given by
\begin{eqnarray}
  \label{eq:average}
  \langle O[\rho] \rangle
& =  & \int\! D[\rho]\ 
  \exp\left( 
    -\hspace{-.3cm}\int\limits_\mathrm{target}\hspace{-.35cm} d x_-
    d^2 x_\perp \frac{\rho^2}{2\mu^2} 
  \right) O[\rho]
\nonumber \\
& = & \int\! D[A_+]\ \exp\left( 
  -\hspace{-.3cm}\int\limits_\mathrm{target}\hspace{-.35cm} d x_-
  d^2 x_\perp 
  \frac{(\nabla_\perp^2 A_+)^2}{2g^2\mu^2} 
  \right) O[\nabla_\perp^2 A_+]\,,
\end{eqnarray}
where the parameter $\mu^2\sim\Lambda^3$ determines the typical
value of the colour charge density. The above average implies a
correlation function
\begin{equation}
  \label{eq:Lambdacorr}
  \langle A_+^a(x_-,x_\perp) A_+^b(y_-,y_\perp)
  \rangle =
  g^2 \mu^2 \delta^{a b} 
  \delta(x_- -y_-) \gamma(x_\perp-y_\perp)
\end{equation}
within the target. Here $\gamma$ is the inverse of the differential
operator in Eq.~(\ref{eq:average}) and is given by
\begin{equation}
  \label{eq:gamma}
  \gamma(x_\perp) = \frac{1}{\nabla^4_\perp} (x_\perp)
  = \int_{\Lambda}\frac{d^2k_\perp}{(2\pi)^2}\,
  \frac{e^{-ik_\perp x_\perp}}{(k_\perp^2)^2}\,,
\end{equation}
where an infrared cutoff $\Lambda$ has been introduced to define
the $k_\perp$ integral. It follows that, at sufficiently small
$x_\perp$,
\begin{equation}
\gamma(x_\perp)-\gamma(0)\simeq\frac{x_\perp^2}{8\pi}\ln(x_\perp^2
\Lambda^2)\,.
\end{equation}
Non-logarithmic terms depend on the specific way of introducing the
cutoff and have been neglected in the above equation.

Using Eqs.~(\ref{eq:average}) and (\ref{eq:Lambdacorr}), the required
average of a product of eikonal factors can be evaluated along the same
lines as the gluon distribution in~\cite{jkmw},
\begin{equation}
  \label{eq:sigmagauss}
  \left\langle\mathrm{tr}\left[U(x_\perp)U^\dagger(x_\perp+\varrho_\perp)
  \right]\right\rangle_{\eta/\Lambda}
  = N_c \exp\left(g^4 C_F  \bar\mu^2 
   \left[ \gamma(\varrho_\perp)-\gamma(0)\right] \right)
\end{equation}
where $\bar \mu^2 = \int\limits_\mathrm{target}\hspace{-.25cm} d x_-
\mu^2\sim\eta\Lambda^2$ reflects the large linear extension $\eta$ of
the target. According to Eq.~(\ref{sc}), an $x_\perp$ integration has
to be performed over a transverse area of a size
$\sim\eta^2/\Lambda^2$, giving rise to the formula
\begin{equation}
\sigma(\varrho)\sim\frac{\eta^2}{\Lambda^2}N_c\left(1-\mbox{exp}\left[
-g^4\frac{C_F}{8\pi}\eta\Lambda^2\varrho^2\ln(1/\varrho^2
\Lambda^2)\right]\right)\,.
\end{equation}
Due to the simple model for $\gamma$ used above, this formula is only
valid for $\rho\ll 1/\Lambda$. However, as has been discussed in
Sect.~\ref{trf}, this is sufficient to determine the virtual photon
cross section for large $\eta$. The result shows the same Glauber type
$(1-\exp)$ structure as Eq.~(\ref{m1}) and the gluon distributions
calculated in \cite{jkmw,km}.

The present analysis is concerned with an energetic colour dipole
created far way from the target and scattering off the target colour
fields. It is the combination of the propagation of this dipole, as
encoded in the Bessel functions in Eq.(\ref{wl}) or the Feynman
propagators of Eq.~(\ref{prop}), and its interaction with the target
that determines the quark distribution. A similar analysis could be
performed replacing the pair of quarks with a pair of gluons. This
would result in a contribution to the gluon density beyond the
instantaneous interactions considered in \cite{jkmw}.

\section{Conclusions}\label{conc}
The scattering of a virtual photon off a large hadronic target, such as
a very large nucleus, has been considered in the high energy limit. The
process has been discussed using both the semiclassical target rest
frame calculation and the McLerran-Venugopalan approach in the infinite
momentum frame. Both approaches show that, for sufficiently large
targets, the cross section is dominated by perturbative contributions.

More specifically, the cross section can be described in terms of the
$q\bar{q}$ component of the photon scattering off the target. For large
targets, the process receives no leading order contributions from quark
pairs with large transverse size. This means that higher Fock states of
the photon are expected to be suppressed by powers of $\alpha_s$ and
the semiclassical approach can be justified in QCD.

One of the main problems of the presented approach is certainly the
procedure of determining the target colour fields which form the
non-perturbative input of the calculation. Two simple models for the
averaging over these fields have been outlined but no satisfactory QCD 
based procedure is presently known.

\setlength{\baselineskip}{.95\baselineskip}

\vfill 
\noindent
{\em We would like to thank W.~Buchm\"uller, A.~Kovner, L.~McLerran
and A.H.~Mueller for helpful discussions and comments.}


\begin{thebibliography}{99}


\bibitem{mv}    L. McLerran and R. Venugopalan, Phys. Rev. D49 (1994) 2233
  
\bibitem{JKWev} J. Jalilian Marian, A. Kovner, A.~Leonidov and H. Weigert
  Nucl. Phys. B504 (1997) 415 and Cavendish-HEP-97/09 (hep-ph/9706377)\\
 J. Jalilian Marian, A. Kovner  and H. Weigert
  Cavendish-HEP-97/15 (hep-ph/9709432)

\bibitem{bk}    J.D. Bjorken and J.B. Kogut, Phys. Rev. D8 (1973) 1341

\bibitem{nz}    N.N. Nikolaev and B.G. Zakharov, Z. Phys. C49 (1991) 607

\bibitem{nac}   O. Nachtmann, Ann. Phys. 209 (1991) 436

\bibitem{bal}   I. Balitsky, Nucl. Phys. B463 (1996) 99

\bibitem{bhm}   W. Buchm\"uller and A. Hebecker, Nucl. Phys. B476 (1996)
                203;\\
                W. Buchm\"uller, M.F. McDermott and A. Hebecker, Nucl. Phys. 
                B487 (1997) 283

\bibitem{bd}    S.J. Brodsky and V. Del Duca, Phys. Rev. D46 (1992) 931

\bibitem{mv1}   L. McLerran and R. Venugopalan, Phys. Rev. D50 (1994) 2225

\bibitem{jaffe} R.L. Jaffe, Nucl. Phys. B229 (1983) 205

\bibitem{bhq}   S.J. Brodsky, A. Hebecker and E. Quack, Phys. Rev. D55 (1997)
                2584

\bibitem{kovchegov} Y.V.Kovchegov, Phys. Rev. D54 (1996) 5463

\bibitem{jkmw}  J. Jalilian-Marian, A. Kovner, L. McLerran and H. Weigert,
                Phys. Rev. D55 (1997) 5414

\bibitem{km}    Y.V.Kovchegov and A.H. Mueller, peprint CU-TP-876
                (hep-ph/9802440)

\end{thebibliography}
\end{document}